\documentclass[11pt]{article}
\usepackage{moriond_alamos,epsfig}

\include{babarsym}

\def\be{\begin{equation}}
\def\ee{\end{equation}}
\def\bea{\begin{eqnarray}}
\def\eea{\end{eqnarray}}

\def\gamtot1s{$\Gamma^{tot}_{\eta_c}$}

\def\babar{\mbox{\sl B\hspace{-0.4em} {\scriptsize\sl A}\hspace{-0.4em} B\hspace{-0.4em} {\scriptsize\sl A\hspace{-0.1em}R}}} 

\def\babarbf{\mbox{\sl \bf B\hspace{-0.4em} {\scriptsize\sl\bf A}\hspace{-0.4em} B\hspace{-0.4em} {\scriptsize\sl\bf A\hspace{-0.1em}R}}} 

\begin{document}
\vspace*{4cm}
\title{Recent results on two-photon and tau physics at \babarbf}

\author{ G. Wagner}

\address{Universit\"at Rostock, Germany, \\
for the \babar\, collaboration}

\maketitle\abstracts{ We present a measurement of the total width
  $\Gamma_{tot}^{\eta_c}$ and mass $M_{\eta_c}$ of the $\eta_c$ meson
  with the \babar\, detector at the PEP-II $B$-factory at SLAC.  The
  results based on a data sample of 88~$fb^{-1}$ are
  $\Gamma_{tot}^{\eta_c} = (33.3\pm2.5(stat.)\pm0.8(sys.)) \mevcc $ and
  $M_{\eta_c}=(2983.3\pm1.2\pm1.8)\,\mevcc $. An enhancement observed around
  3.63 $\gevcc$ can be interpreted as an evidence for the $\eta_c(2S)$
  state. In addition a preliminary measurement of the $\tau$ lepton
  lifetime using 30.2 $fb^{-1}$ is described. The measured $\tau$
  lifetime is: $\tau_{\tau} = (290.8\pm1.5(stat.)\pm1.6(syst.)) \fs$. }

\section{Measurement of the $\eta_c$ meson mass and width}
\subsection{Introduction}
The mass and the width of the charmonium ground state $\eta_c(1S)$ are
not well established. The world average~\cite{PDG2002} of the total
width and mass is $\Gamma_{tot}^{\eta_c} = 16.0^{+3.6}_{-3.2}\,\mevcc
$ and $M_{\eta_c} = (2979\pm1.5)\, \mevcc$, respectively.  The $\eta_c$ is expected
to decay dominantly via two-gluon annihilation, so that
$\Gamma_{tot}^{\eta_c}\approx \Gamma_{gg}^{\eta_c}$. Hence the ratio
of the total width of the $\eta_c$ to its two-photon partial width is
predicted by NLO perturbative QCD~\cite{kwong}:
\begin{equation}
\frac{\Gamma^{\eta_c}_{tot}}{\Gamma^{\eta_c}_{\gamma\gamma}}\approx\frac{9\alpha_s^2}{8\alpha^2}\cdot\frac{1+4.8\alpha_s/\pi}{1-3.4\alpha_s/\pi}
\end{equation}
Using~\cite{kwong} $\alpha_s = 0.28\pm0.02$ and the world
average~\cite{PDG2002} of $\Gamma^{\eta_c}_{\gamma\gamma} =
7.5\pm0.8\, \kev$ gives $\Gamma^{\eta_c}_{tot}$ as $25.4\pm6.0\,\mevcc
$, a value higher than but still consistent with the world average.

\subsection{Fit of the mass spectrum}

The data sample consists of an integrated $e^+e^-$ luminosity of 88
$fb^{-1}$, including runs on and below the $\Upsilon(4S)$ resonance. The
$\eta_c$ mesons are produced by two-photon interactions. They are
selected via the decay channel $\eta_c \rightarrow \KS
K^{\pm}\pi^{\mp}$ with $\KS \rightarrow\pi^+\pi^-$, where in addition
the $e^-$ and $e^+$ are required to escape through the beam pipe.

The mass spectrum is shown in figure \ref{fig::etac}. A large peak can
be seen at the $\eta_c$ mass, a smaller one at the $J/\psi$ mass.
$J/\psi$ are produced by $e^+e^-$ annihilation where one electron
emitted a hard photon by initial state radiation.

The width of the $\eta_c$ is of the same order of magnitude as the
mass resolution of the detector. Therefore the detector resolution has
to be precisely known in order to extract the $\eta_c$ width from the
observed peak in the invariant mass spectrum. The width~\cite{PDG2002} of the
$J/\psi$ of $87\pm5\, \kev$ is much smaller than the
detector resolution. Hence the observed width of the $J/\psi$ peak is
completely determined by the resolution.

The mass spectrum is fitted in the mass range from 2.4 to 3.6
$\gevcc$. The $\eta_c$ is represented by a convolution of a
non-relativistic Breit-Wigner function and a Gaussian. The $J/\psi$
peak is fitted with a single Gaussian. The standard deviation
$\sigma_{\eta_c}$ of the Gaussian, which describes the mass resolution
at the $\eta_c$ peak, is constrained to a value 0.8 $\mevcc $ lower
than that for the $J/\psi$. This difference of 0.8 $\mevcc $ of the
mass resolution was obtained from Monte Carlo events.  The background
is represented by the function: $A\cdot e^{-B\cdot mass}$.  The free
fit parameters are the $J/\psi$ mass $M_{J/\psi}$, the mass
difference $M_{\eta_c} - M_{J/\psi}$, the $\eta_c$ width
$\Gamma_{tot}^{\eta_c}$, the mass resolution of the $J/\psi$
$\sigma_{J/\psi}$, the normalization and slope of the background and
the number of events in the $\eta_c$ and $J/\psi$ peaks. The result of
the fit is presented in table \ref{tab::etac}.

\begin{table}[!h]
\caption{Result of the unbinned maximum likelihood fit of the mass spectrum. The resolutions of the $J/\psi$ and the $\eta_c$ peaks are $\sigma_{J/\psi}$ and $\sigma_{\eta_c}$, respectively \label{tab::etac}.}
\vspace{0.4cm}
\begin{center}
\begin{tabular}{|c|c|c|c|c|c|}
\hline
 & $\Gamma_{tot}^{\eta_c}$ & $M_{J/\psi}-M_{\eta_c}$ & $M_{J/\psi}$ & $\sigma_{J/\psi}$ & $\sigma_{\eta_c}$ \\ 
 & $[\mevcc ]$ &$[\mevcc ]$ & $[\mevcc ]$ & $[\mevcc ]$ & $[\mevcc ]$ \\ \hline
 fit & $33.3\pm2.5$ & $113.6\pm1.2$ & $3094.0\pm0.8$ & $7.5\pm0.8$ & $\sigma_{J/\psi} - 0.8$ \\ \hline
\end{tabular}
\end{center}
\end{table}

\vspace{-0.5cm}

\subsection{Systematic errors}

A source of a systematic error on $\Gamma_{tot}^{\eta_c}$ is the
uncertainty of the detector mass resolution. This contribution is
estimated to 0.4 $\mevcc $ by using the mass resolution for the
$\eta_c$ obtained from Monte Carlo events for the fit. The systematic
error due to the parametrization of the background is estimated by
varying the mass range of the fit.  A variation from 2.4-3.6 $\gevcc$
to 2.7-3.3 $\gevcc$ changes the value of $\Gamma_{tot}^{\eta_c}$ by
0.7 $\mevcc $.

The reconstructed masses $M_{J/\psi}$ and $M_{\eta_c}$ are shifted in
Monte Carlo simulation by -1.1~$\mevcc $ compared to the values used
as input.  This bias does not affect the mass difference. However the
fitted $J/\psi$ mass is still shifted by -1.8 $\mevcc $ from the well
established PDG value~\cite{PDG2002} of $(3096.87\pm0.04) \mevcc$
after correction of the bias observed in the Monte Carlo simulation.
We take this shift as a systematic error on the $\eta_c$ mass since we
do not know its source. The unkown mass shift is not necessarily the
same for the $\eta_c$ and the $J/\psi$ because the angular
distribution of their decay products are different.  The shift could
be caused by inhomogeneities in the magnetic field which may not be
correctly modeled in Monte Carlo.

\subsection{Evidence for the $\eta_c(2S)$ state}

The $\eta_c(2S)$ is not experimentally established yet~\cite{PDG2002}.
Its theoretically predicted mass is in the range~\cite{etac2sTheory}
3583 to 3640 $\mevcc $. There exist two experimental
results~\cite{CrysBall}$^,$ \cite{BelleEtac2s} which are not in agreement.

\begin{figure}[!h]
\begin{center}
 \psfig{figure=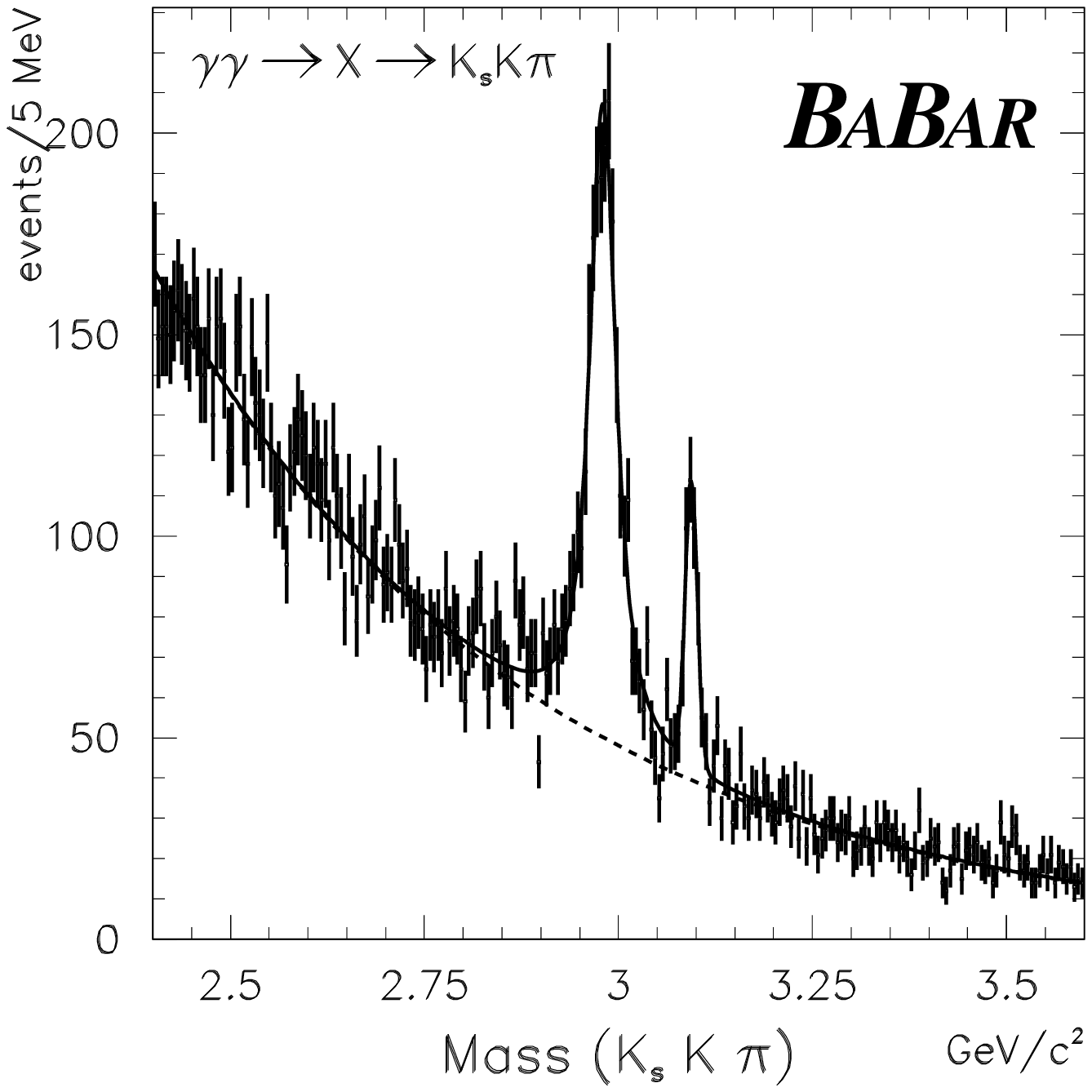,height=7.0cm}
 \psfig{figure=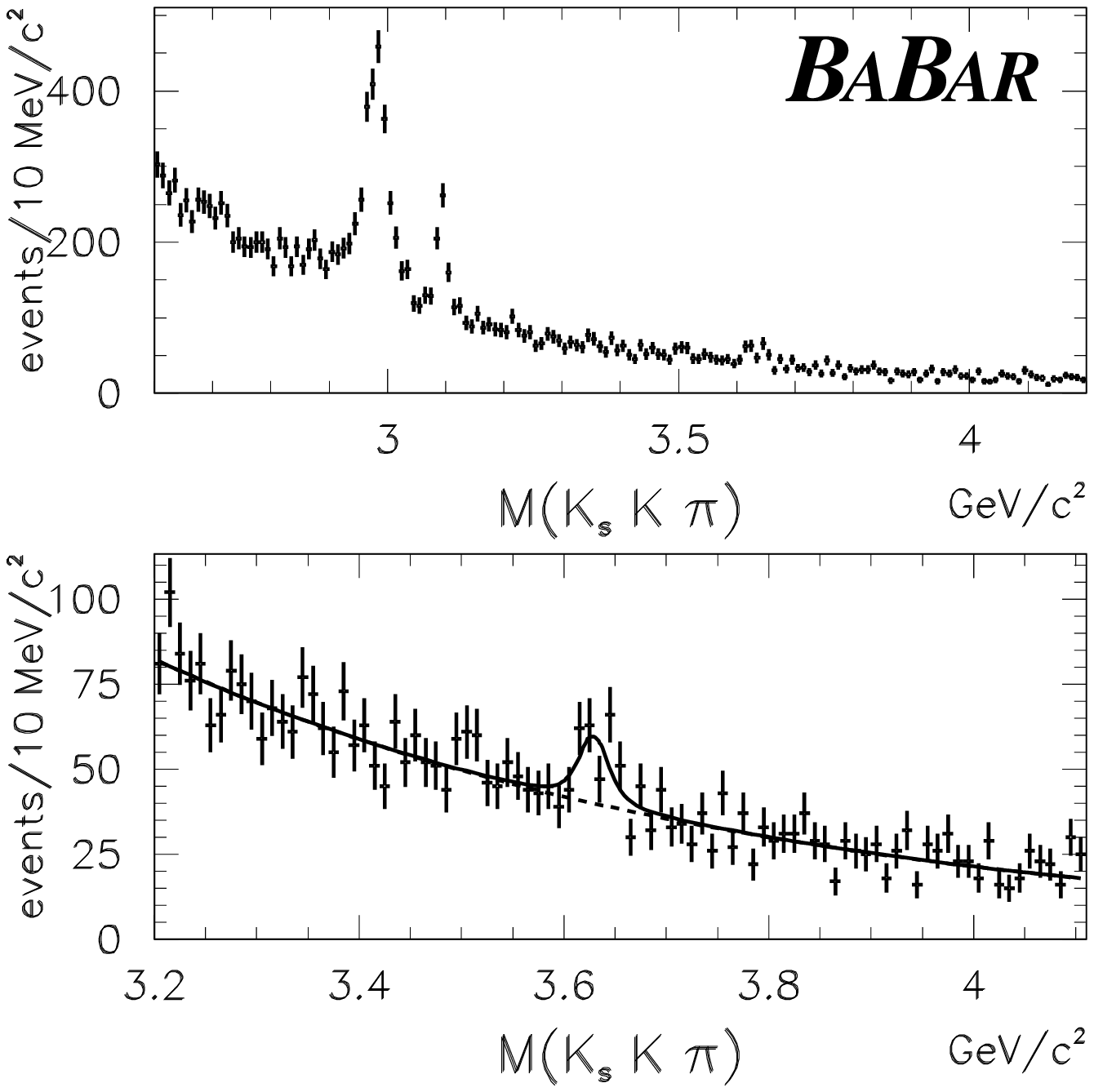,height=7.0cm}
\caption{Various histograms of the $\KS K^\pm\pi^\mp$ invariant mass spectrum. The large peak around 2983 $\mevcc $ is the $\eta_c$, the peak next to it is the $J/\psi$. An evidence of a $\eta_c(2S)$ signal can be seen around 3630 $\mevcc $. \label{fig::etac}}
\end{center}
\end{figure}

The upper right histogram of figure \ref{fig::etac} shows the $\KS
K^{\pm}\pi^{\mp}$ invariant mass spectrum extended to higher masses. An
enhancement can be seen at about 3.63 $\gevcc$ which lies within the
predicted mass region of the $\eta_c(2S)$ and hence is interpreted as
an evidence of this state. The mass spectrum is fitted using a
Breit-Wigner function convoluted with a Gaussian to describe the
enhancement. The standard deviation of the Gaussian is set to 8.5
$\mevcc $ which was estimated from a Monte Carlo sample.  The fit
results in $\Gamma_{tot}^{\eta_c(2S)} = (20\pm10)\, \mevcc $ and
$M_{\eta_c(2S)} = (3633.3\pm5.0)\,\mevcc $. The $\eta_c(2S)$ peak
contains $86\pm23$ events.

The systematic error associated to the $\eta_c(2S)$ width due to an
uncertainty of the mass resolution is 1.6 $\mevcc $. The background
parametrization gives a contribution of 3.3 $\mevcc $ to the systematic
error on the total width.

In summary, we have measured the mass and width of the $\eta_c$ meson 
$M_{\eta_c} = (2983.3 \pm 1.2\,(stat)\, \pm 1.8\,(syst))\, \mevcc $
and $\Gamma_{tot}^{\eta_c} = (33.3 \pm 2.5\,(stat)\,\pm
0.8\,(syst))\,\mevcc $ and observed an evidence of the $\eta_c(2S)$.
The mass and width of the $\eta_c(2S)$ were measured to $
M_{\eta_c(2S)} = (3632.2 \pm 5.0 \pm 1.8)\, \mevcc $ and
$\Gamma_{tot}^{\eta_c(2S)} = (20\pm10\pm4)\, \mevcc $, respectively.

The measured value of the total width of the $\eta_c$ deviates from
the world average by more than three standard deviations while the
measured mass agrees well with the world average. The mass of the
enhancement interpreted as the $\eta_c(2S)$ falls in the
theoretically predicted range of the $\eta_c(2S)$ mass.

\section{Measurement of the $\tau$ lepton lifetime}

A measurement of the $\tau$ lepton lifetime, combined with the
branching ratio of $\tau$ into leptons, the $\tau$ and $\mu$ mass and
the $\mu$ lifetime, provides a test of lepton universality. This test
is experimentally limited by the uncertainties of the $\tau$ lifetime
and its leptonic branching fraction.

The mean $\tau$ lifetime $\langle \tau_{\tau} \rangle$  is given by 
\begin{equation}
\langle \tau_{\tau} \rangle = \frac{M_{\tau}}{\langle P_{\tau} \rangle} \langle \lambda_{\tau} \rangle, 
\end{equation}
where $M_{\tau}$ is the $\tau$ mass, $\langle P_{\tau} \rangle$ the
average $\tau$ momentum in the center-of-mass frame and $\langle
\lambda_{\tau} \rangle$ the mean $\tau$ decay length. The average
momentum is calculated from a Monte Carlo sample. The mean decay length is
measured using decays where one $\tau$ decays into three charged
particles while the other one decays into a single charged particle.

\begin{figure}[!h]
\begin{center}
\parbox{7.5cm}{\psfig{figure=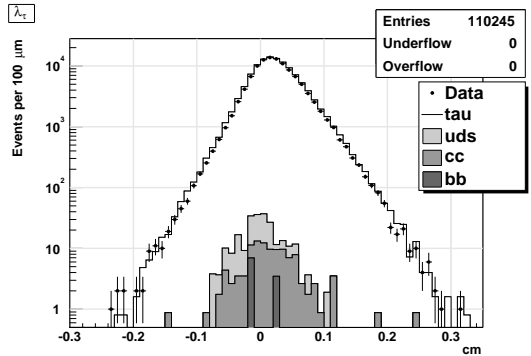,height=4.8cm}}
\parbox{5.5cm}{\caption{Measured decay length for data and Monte Carlo. \label{fig::tau} }}
\end{center}
\end{figure}

The tracks of the 3-prong $\tau$ candidate are projected onto the
transverse plane, which is perpendicular to the boost direction of the
$e^+e^-$ annihilation. The 3-prong vertex $\vec{x}_{3p}$
in the transverse plane is reconstructed and the transverse decay length
$\lambda_{\tau,t}$ is calculated by:
\begin{equation}
\lambda_{\tau,t} = (\vec{x}_{3p}-\vec{x}_{bs})\cdot \hat{p}_{3p}
\end{equation}
where $\vec{x}_{bs}$ is the center of the luminous region and
$\hat{p}_{3p}$ the unit vector of the 3-prong momentum in the
transverse plane.  This method of reconstructing the decay length was
chosen because it is less dependent on uncertainties of the detector
alignment. The decay length in space $\lambda_{\tau}$ is calculated by
dividing the transverse decay length by the sine of the polar angle of
the 3-prong momentum in the center-of-mass system.  Figure
\ref{fig::tau} shows the decay length distribution for $\tau$
candidates.

The mean decay length is calculated by averaging over all events.
The averaging is done without any error weighting of the events.
Instead an azimuthal weighting is done to achieve an uniform azimuthal
event distribution. This method of averaging is more robust with
regards to uncertainties of the position of the luminous
region.

\vspace{-0.3cm}

The bias of this analysis method was obtained from a Monte Carlo
sample. Its uncertainties are the main contribution (1.3 fs) to the
systematic error of the $\tau$ lifetime. The detector alignment is 
another major source of systematic uncertainties (0.7 fs).

The result based on $30.2\,fb^{-1}$ is $\tau_{\tau} =
(290.8\pm1.5(stat.)\pm1.6(syst.)) \fs$. It agrees well with the
world average of $(290.6\pm1.1) \fs$.

\section*{Acknowledgments}

We are grateful for the excellent luminosity and machine conditions
provided by our \pep2\ colleagues, and for the substantial dedicated
effort from the computing organizations that support \babar.  The
collaborating institutions wish to thank SLAC for its support and kind
hospitality.  This work is supported by DOE and NSF (USA), NSERC
(Canada), IHEP (China), CEA and CNRS-IN2P3 (France), BMBF and DFG
(Germany), INFN (Italy), FOM (The Netherlands), NFR (Norway), MIST
(Russia), and PPARC (United Kingdom).  Individuals have received
support from the A.~P.~Sloan Foundation, Research Corporation, and
Alexander von Humboldt Foundation. The author would like to thank the
organizers of the {\em XXXVIIIth Recontres de Moriond} for a wonderful conference.

\end{document}